\documentstyle[sprocl,psfig]{article}

\def\be{\begin{equation}}
\def\ee{\end{equation}}
\def\bea{\begin{eqnarray}}
\def\eea{\end{eqnarray}}

\begin{document}

\title{Physics for a Polarized Electron-Nucleon/Nucleus-Collider
\footnote{Invited talk at EPIC 99; April 8-11, 1999, Indiana University
  Cyclotron Facility,\\ Bloomington, USA}}

\author{A. Sch\"afer}

\address{Institut f\"ur Theoretische Physik\\
University of Regensburg\\ D-93040 Regensburg\\ GERMANY\\
E-mail: andreas1.schaefer@physik.uni-regensburg.de} 


\maketitle\abstracts{ 
A number  of physics arguments for a high-luminosity 
high energy  polarized Electron-
Nucleon/Nucleus-Collider 
(i.e. with a luminosity of at least $10^{33}$ cm$^{-2}$ sec$^{-1}$
and an invariant energy squarred of at least  $s\geq 100$ GeV$^2$)
are presented. The main purpose of
this machine would be generally speaking the investigation of QCD 
beyond the twist-2 level respectively of nuclear  physics beyond 
the level of 'effective' models. Specific topics are: twist-2 
and
twist-3 spin-asymmetries as probes of both the internal hadron
structure and the hadronization process,
Off-Forward-Parton-Distributions as a new dimension of QCD-physics, 
nuclear effects for the nucleon structure and nuclear effects for
QCD-dynamics in nuclei. The last two topics are also of direct
relevance for high-energy heavy-ion-collisons. We conclude that the
need for such a collider is clear, if nuclear physics is to continue
its development towards a comprehensive understanding of QCD phenomena.}

\section{Introduction}
In recent years a marked development took place in which the hot
topics of QCD moved more and more into the reach of the  traditional 
nuclear physics community. This is illustrated most clearly by 
high-energy heavy-ion physics ({\sc rhic} and {\sc alice}) but 
is also visible in the fact that many members  of 
{\sc hermes}, {\sc compass}, the {\sc slac}-spin-collaboration, and other
similar collaborations
have  a nuclear physics background. Physicswise this development is 
fueled by the fact that most of the recent important progress in 
QCD concern higher-twist-effects, specific hadron-wavefunctions,
low-energy limits of QCD, quasi-classical 
(Glauber-type) approximations for high-energy heavy-ion physics, ...,
all of which expand the scope of QCD into the direction of  
more traditional nuclear physics. Consequently for the first time 
there is well-founded hope to finally be able to join both
descriptions in a consistent manner. The main task for the next
decade(s) as seen by many 
theoreticians and experimentalists is therefore to  
close the remaining gap in our understanding. Consequently, the
discussions with respect to the most suitable machine 
for this purpose became rather lifely in recent years.
Of these ideas those which are closest to the specific machine 
we are discussing here are:  to add an electron-ring to {\sc rhic},
to more or less rebuildt {\sc cebaf} as a 25 GeV machine, to performe
high luminosity experiments with an {\sc elfe} detector either at {\sc
  desy} or at {\sc cern}, etc.
The main task is  therefore to clarify, which physics questions favour
which of these machines. In the following I shall address some 
of the relevant research topics, and shall trying to provide some 
(very partial) answers.\\

\section{Higher-Twist versus Higher-Order-Perturbation-Theory }

Many of the presently most hotly discussed issues in QCD focus on 
higher-twist-phenomena. To the extent that the understanding increases
the meaning of this term in different contexts becomes, however, more 
and more spezialised and thus ever more confusing to non-experts.
Let me try to illustrate this with a few remarks which at the same time
sketch the vast field of research for an {\sc epic}-type collider.
Originally 'higher-twist' was a rather general term for all
processes suppressed by some hard scale, i.e. by powers of $1/Q^2$ for
deep-inelastic scattering. This made sense in times when one was only 
aiming at finding some arguments to neglect these terms. By now one is
primarily interested to understand their origin and such a broad 
definition is not really helpful. In a first step it is certainly
useful  to distinguish between 'power-corrections' like e.g. rather
trivial finite-mass-corrections and terms which are due to
correlators of more than two quark or gluon fields. The latter 
can, however, again be of different nature:\\

{\bf Higher-Twist in the framework of Operator-Product-Expansion}
The term 'higher-twist' was originally defined within OPE. Here the moments
of structure functions can be represented by specific 
local correlators of 
a certain number of fields within the hadron under investigation. A
well
known example is the third moment of the second spin structure
function
$g_2(x,Q^2)$, which is given by \cite{AS1} 
\be
\int_0^1 g_2^{p,n}(x,Q^2) x^2 dx ~=~ -{1\over 3} a_2 + {1\over 3} d_2
+ ...
\ee
\be
d_2 \approx \langle PS \| \bar q \gamma_{\mu} \tilde G_{\alpha\beta}
\psi \| PS\rangle
\ee

\begin{figure}[htb]
\centerline{\mbox{\psfig{file=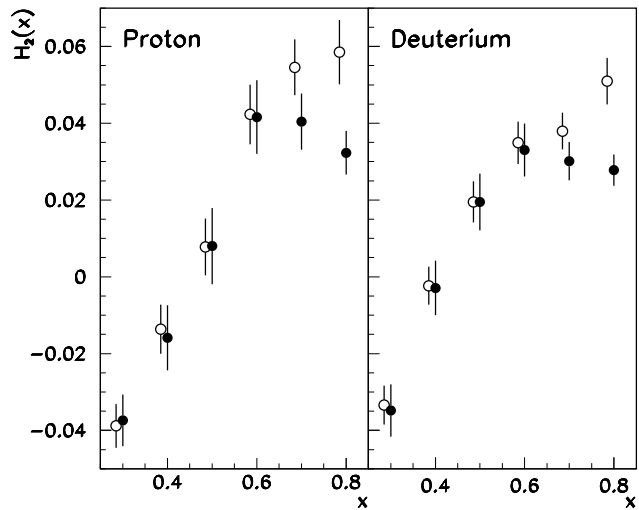,width=0.7\linewidth}}}
\bigskip
\caption{Genuine higher twist contributions as extracted from 
the presently available data on $F_2(x,Q^2)$. ($H_2(x)=\left( 
F_2^{\rm higher~~ twist}(x,Q^2) - F_2^{\rm target~~mass~~ 
corrections}(x,Q^2)\right) 
{Q^2\over 1 ~{\rm GeV}^2}$) }
\label{fig1}
\end{figure}

\begin{figure}[htb]
\centerline{\mbox{\psfig{file=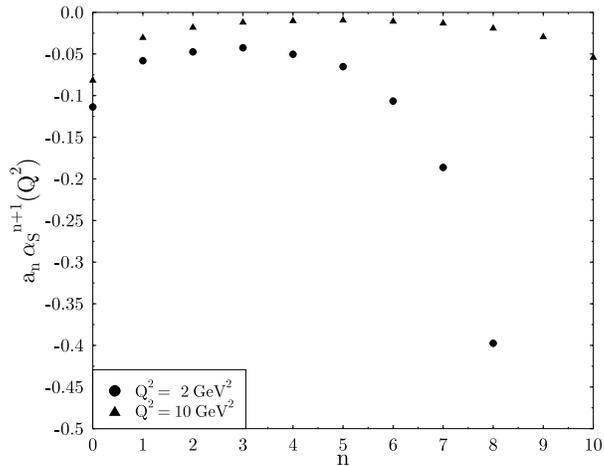,width=0.7\linewidth}}}
\bigskip
\caption{Perturbative corrections to Bjorken sum rule calculated in
  the Naive-Nonabelianization-Approximation, for two different values of
  $Q^2$. The figure illustrates, that perturbative corrections 
define  only an  asymptotic series, that for medium large $Q^2$ they
start to diverge already in rather low order, and that the convergence
improves with increasing $Q^2$.}
\label{fig2}
\end{figure}

To distinguish higher twist from leading twist contributions
by simply fitting the $Q^2$ dependence of structure functions 
requires very high statistics and a large kinematic domain 
(which in turn requires a sufficently high $s$).
This is so
far only possible for the unpolarized case, see e.g.  \cite{HT1}
and figure \ref{fig1}, taken from that paper. The values plotted in
this figure are the fitted $1/Q^2$-corrections minus the target mass 
corrections 
\be
H_2(x)=\left( F_2^{\rm higher~~ twist}(x,Q^2) - F_2^{\rm target~~mass~~
corrections}(x,Q^2)\right) {Q^2\over 1 ~{\rm GeV}^2}
\ee
As the leading twist contribution to the third moment of 
$g_2$ is supressed, it is also 
possible to extract $d_2$ from these data  even for very
limited statistics \cite{E154}. A far more precise determination will
soon be published by {\sc e-155x}. The determination of
higher-twist-correlators from moments of structure functions 
does not only provide well defined 
information on the internal nucleon wave function which in turn 
allows to exclude many  models
and puts the different  calculation techniques to a crucial test.
In addition it also 
addresses a fundamental problem of quantum field
theory in general. Perturbative expansions are only asymptotic series,
i.e. they blow up at some order in $\alpha_s$.
This is illustrated in figure \ref{fig2}
for the perturbative corrections to Bjorken sum rule as calculated
in 'Naive-Nonabelianization-Approximation', i.e. iterating only those
contributions contained in the leading order $beta$-function.
\be
\int_0^1dx (g_1^p(x)-g_1^n(x)) ~=~ {g_A\over 6g_V} \left(
1+\sum_{n=0}^{\infty} a_n \alpha_s^{n+1} \right)
\ee
Obviously the convergence of this series is rather bad for the 
typical $Q^2$ values of e.g. the {\sc slac} and 
{\sc hermes} spin experiments.
Only the sum of higher-twist corrections (better to be called
power-corrections in this context) perturbative contributions 
and genuine non-perturbative (like e.g. instanton) contributions gives
the physical result. 
The increase in the higher-order  perturbative contributions is e.g. 
canceled by the power corrections \cite{ren}.
The individual contributions are furthermore  in general scheme
dependent, somewhat comparable to the fact that  in NLO the
distribution 
functions in DIS become scheme dependent. For NLO (and NNLO etc.)
calculations one has learned how to handle this, for the power
corrections a full understanding still has to be developed.
Let us note as an illustration, that the perturbative corrections plus
the renormalon-power-corrections as calculated in the $\overline{MS}$
scheme 
can be reexpressed as just the first order perturbative 
corrections with a suitable redefined scale-parameter 
(BLM-scheme) \cite{BLM}. 
To understand these fascinating fundamental aspects of
quantum field theory in depth one really needs precisely measured 
$Q^2$-dependences of as many moments of structure functions as
possible.\\

\newpage
{\bf Higher-twist in semi-inclusive ractions}\\
The investigation of semi-inclusive reactions of the type
\be
e+p~~ \rightarrow ~~e' ~+~ h_1 ~(+h_2)~ +X
\ee
plays already a major role for the ongoing experiments and will
do much more so for all planned ones. The theoretical analyses
of these reactions is very involved and combines leading and higher
twist distribution and fragmentation functions. 
Reactions of the type (3) are already presently used for many
applications. Let me give just a very few  examples. From pion production
in polarized lepton-nucleon collisions {\sc hermes} and {\sc smc} 
gained information on the flavour decomposition of the nucleon spin,
which is independent from those information contained in
the inclusive structure functions. {\sc hermes} was even able to
obtain a first direct glimps of the polarized gluon distribution,
which will be investigated in far more detail by {\sc compass},
e.g. by using open charm production. Also charm and strangeness
production in unpolarized collisons is of great interest, as it 
helps to clarify how much strange and charm quarks there are in a
nucleon. Recent {\sc hera} data strongly suggests that there could 
e.g. be substantial intrinsic charm in the nucleon (see the talks by 
T. Adams and R.
Vogt). There are actually by now so much data and so many 
theoretical aspects to this kind of reactions, that it is not possible
to review it in any reasonable manner here.
Some more information is given in the talks by Melnitchouk, Radici,
Carlson and de Florian.  \\
It turned out that this type of reactions is full of surprises, such
that there is a constant need for improved experimental data (for
larger kinematic regions and with higher statistics).
As an example let me mention that {\sc hermes} recently observed 
for the first time a specific
azimuthal single spin asymmetry which can be related 
to well defined combinations of up to now completely unknown
distribution and fragmentation functions.
For detailed discussion of the complicated situation 
arising from the inclusion of spin 
please see  the talk by Piet Mulders.
It seems save to conclude that semi-inclusive and (exclusive)
reactions will play the most important role for any future 
project on lepton-nucleon/nucleus scattering. A collider geometry 
and kinematics is extremely helpfull for such studies, as it makes
also the target fragmentation region fully accessible.\\

{\bf Higher-twist wave-functions and exclusive reactions}\\
Also totally exclusive reactions at high $Q^2$ gain more and more 
interest.  The task here is to pin down the hadron-wave functions 
(i.e. the lightcone wave-functions) with increasing precision. This
leads to an expansion which can be thought of as a
kind of Fock-state expansion in a specific kinematics. For the 
$\rho$ meson the quantities of interest are e.g. of the type
\cite{braun}
\be
\langle 0|\bar u(z) ig\tilde G_{\mu\nu}(vz) d(-z) |\rho^-(P,\lambda)
\ee
Expansions keeping the higher order terms which contain more than 
the minimum number of
fields are called higher-twist distribution amplitudes.
Again the meaning of the term is slightly different in this context.
A large fraction of the experiments to be done at an {\sc
  epic}-type
machine will address this type of higher-twist contributions.
It should also be noted that heavy quark physics, which will gain
substantially in 
importance over the next years should accelerate the development in
this field. Many hadron decays involving heavy quarks can be 
analysed in terms of such higher-twist distribution amplitudes.
Their detailed experimental investigation will catalyze their improved
theoretical description (and vize versa). One can therefore forsee
that
by the time an {\sc epic}
type machine would start operating a large number of observables will
have a very well defined significance in terms of clearly defined QCD
amplitudes. On the other hand the expected progress
will lead to substantial demand for complementary experiments. As 
for this type of physics luminosity is more important than high
energy, it is, however, not obvious that an {\sc epic} type machine
would be better suited than an {\sc elfe} type one.\\

\begin{figure}[htb]
\centerline{\mbox{\psfig{file=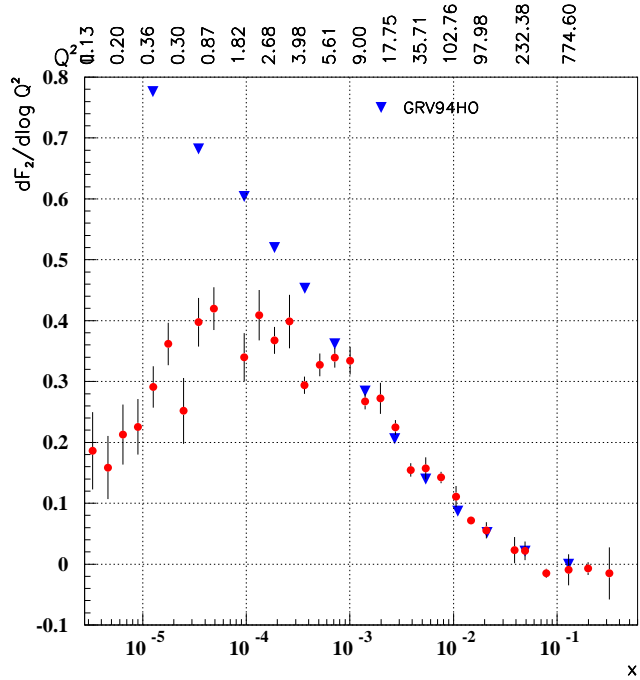,width=0.7\linewidth}}}
\bigskip
\caption{The Caldwell plot for recent {\sc hera} data. The 
logarithmic derivative $d F_2/d \log(Q^2)$ should be
proportional to $xG(x,Q^2)$.}
\label{fig3}
\end{figure}

{\bf Higher-Twist Evolution and saturation in heavy ions}\\
One of the crucial
questions for high-energy heavy-ion-collisions and small-$x$ physics
as investigated e.g. by {\sc hera} is the appearence of non-linear
effects in the $Q^2$-evolution equations at very small $x$. These are
driven by the gluon distribution functions becomming large and should
at the latest be relevant if $\alpha_s G(x,Q^2)$ becomes larger than
one.
It seems that {\sc hera} is just able to touch this region. For nuclei
these non-linear effects should set in much earlier as the soft gluons
becomes delocalized within the line of sight within a nucleus, leading
to an $A^{1/3}$ enhancement factor. As $G(x,Q^2)$ is given by the
square of a gluon-field operator with dimension (energy)$^2$ these
non-linear terms are typically suppressed by a factor 1/$Q^2$.
For a specific example see e.g. \cite{AMu} where the conclusion is
reached that in heavy nuclei the quark and gluon distributions 
behave like 
\bea
{d(xq(x))\over d^2bd^2l}& = {N_c\over 6\pi^4} {Q^2_s\over l^2} 
~~~~~~~~~~{\rm for} ~~~~~~~~~~l^2 \gg Q_s^2 \nonumber \\
& =  {N_c\over 2\pi^4}  
~~~~~~~~~~{\rm for} ~~~~~~~~~~l^2 \ll Q_s^2 
\eea
\bea
{d(xG(x))\over d^2bd^2l}& = {N_c^2-1\over 4\pi^4} {Q^2_s\over l^2} \ln(1/x) 
~~~~~~~~~~{\rm for} ~~~~~~~~~~l^2 \gg Q_s^2 \nonumber \\
& =  {N_c^2-1\over 4\pi^4} \ln(1/x) 
~~~~~~~~~~{\rm for} ~~~~~~~~~~l^2 \ll Q_s^2 ~~~~~~~,
\eea
i.e. the parton distributions are suppressed by a factor $1/Q^2$ when 
the transverse momentum of the outgoing quark $l$ is small  compared
to some saturation momentum $Q_s$. $b$ is the impact parameter of the 
collison. This problem has been approached along many different lines,
some of which seem to be more or less equivalent in comparable
semi-classical approximations \cite{shad}. The characteristic property
shared  by all of them is that 
for very small-$x$  a saturation of the gluon
density is suggested. 
There is even some experimental evidence for such a 
saturation process in recent  {\sc hera} data
\cite{cald}. Figure \ref{fig3} 
shows $d F_2/d \log(Q^2)$, a quantity 
which should in leading order be proportional to the gluon
distribution function $xG(x,Q^2)$, as a function of $x$.
The marked decrease below $x=10^{-4}$ is taken as evidence for
saturation.  
To investigate this specific kind of  higher-twist effects 
in nuclei one would like to have, however, more
energy than envisaged for {\sc epic}. 
Assuming that the nuclear effects compensate about one to two orders
of magnitude in $x$ one finds that with $s=1000$ GeV$^2$ one could 
reach a domain comparable to {\sc hera} and thus start to
see saturation effects (just as in figure \ref{fig3}).
An electron ring at {\sc rhic} 
or heavy ions in {\sc hera} are probably the better 
options for this kind of physics.

\section{Off-Forward-Parton-Distributions (OFPD)}

\begin{figure}[htb]
\centerline{\mbox{\psfig{file=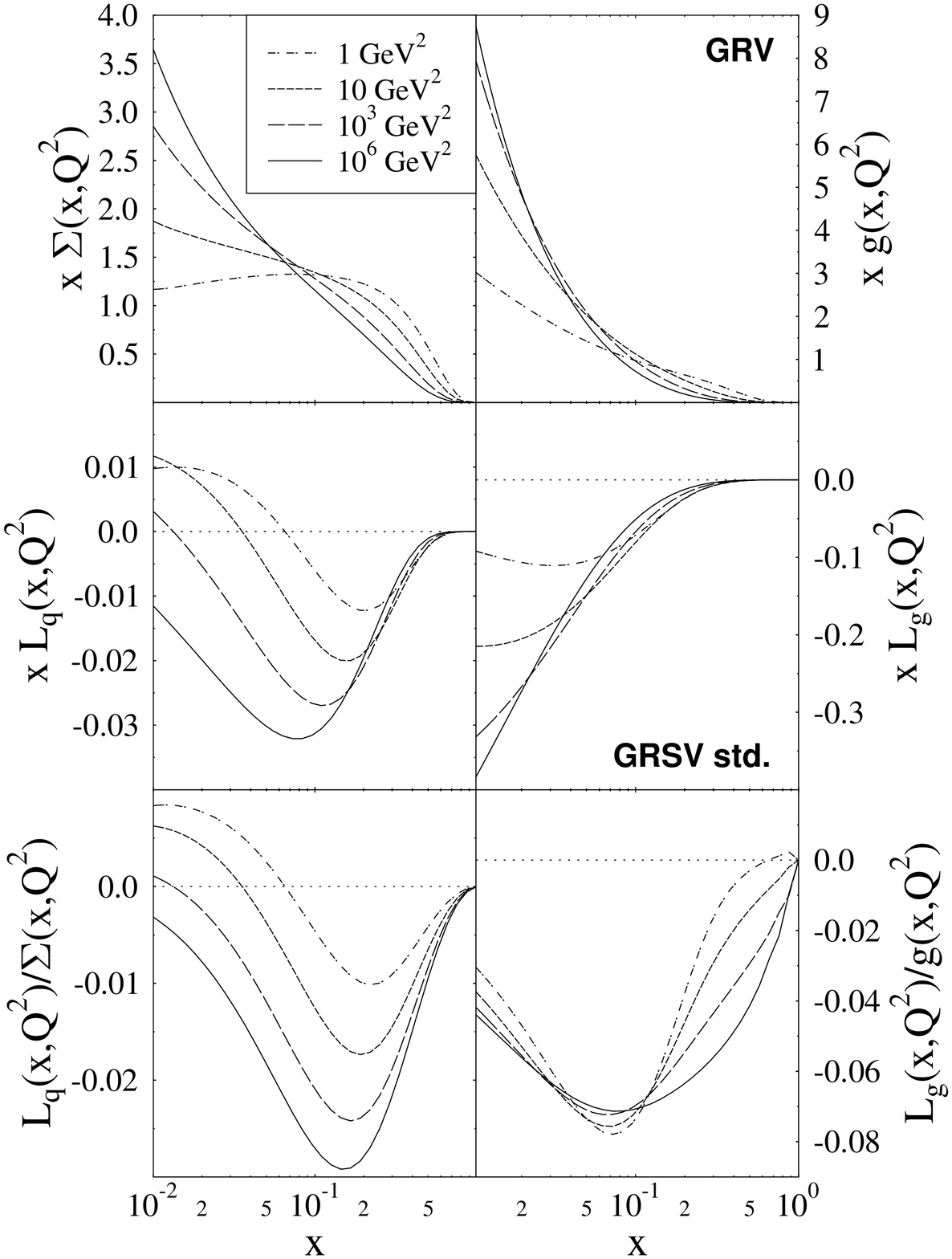,width=0.7\linewidth}}}
\bigskip
\caption{Result of $Q^2$-evolution for the GRV leading order standard
  scenario with $\mu_0^2=0.23$ GeV$^2$}
\label{fig4}
\end{figure}

\begin{figure}[htb]
\centerline{\mbox{\psfig{file=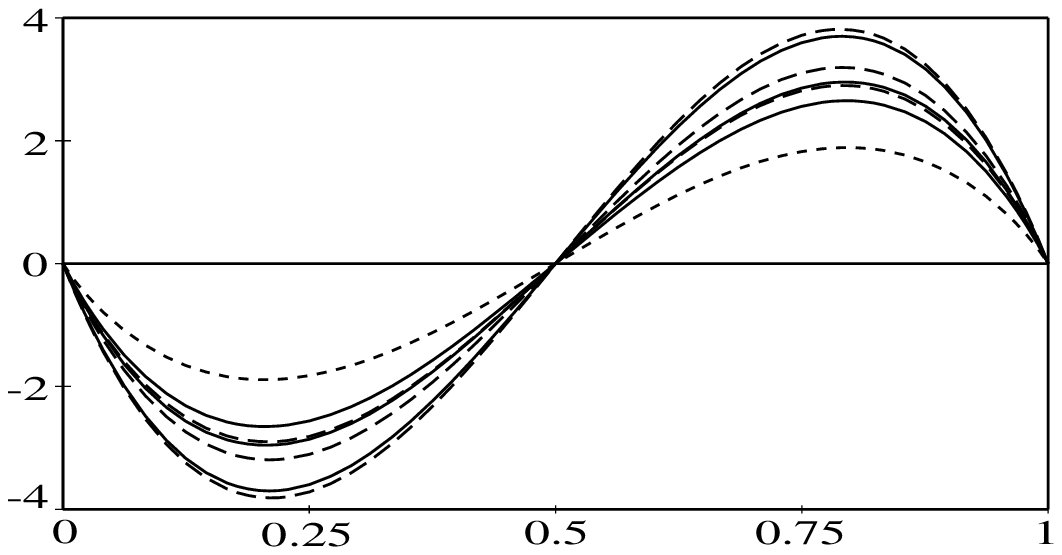,width=0.7\linewidth}}}
\centerline{\mbox{\psfig{file=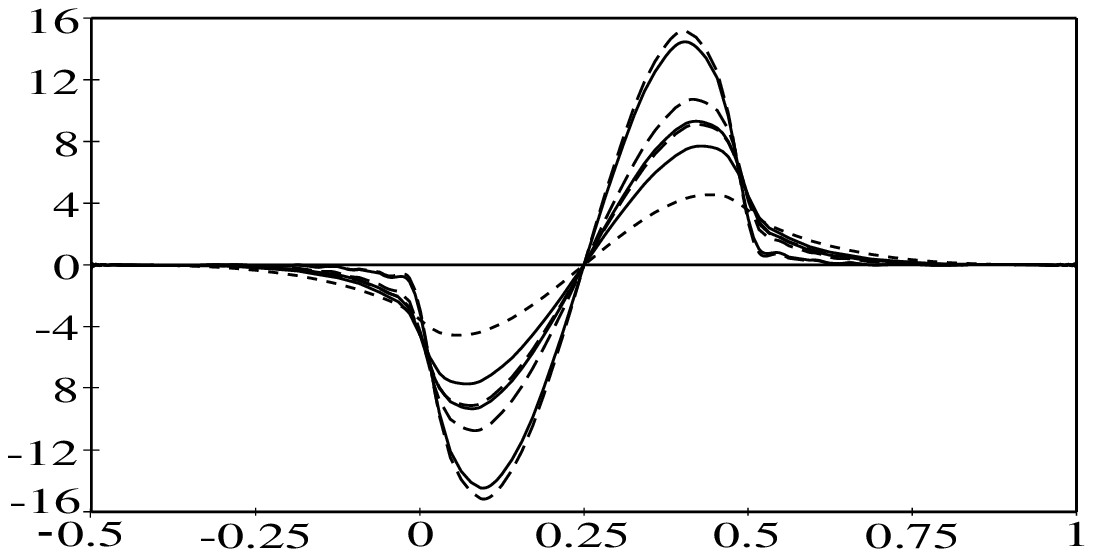,width=0.7\linewidth}}}
\centerline{\mbox{\psfig{file=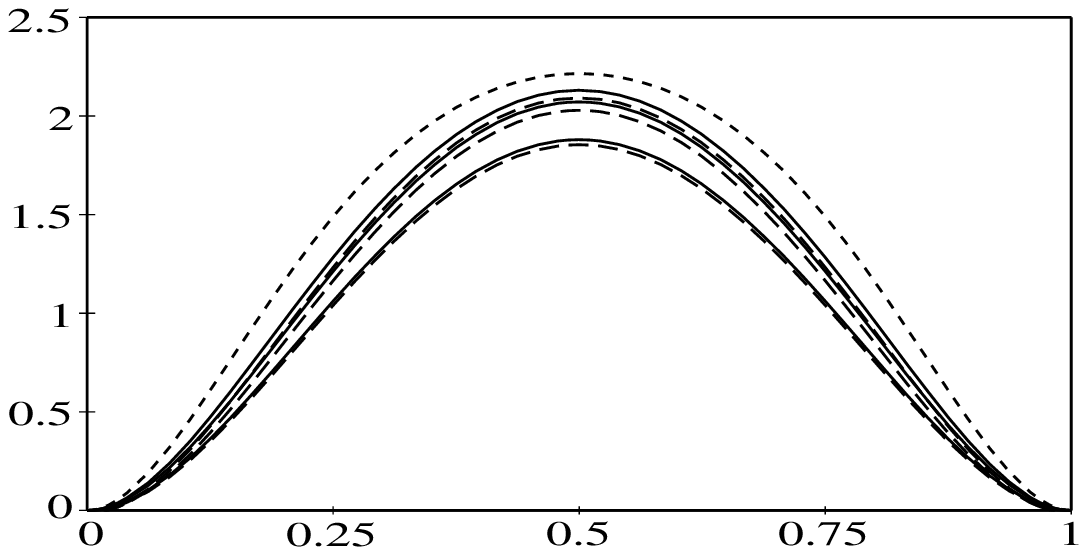,width=0.7\linewidth}}}
\bigskip
\caption{ Evolution of the non-forward singlet quark
distributions ${\cal Q} (x, \zeta)$. The input function at $Q_0 = 0.7\
{\rm GeV}$ is shown by the short-dashed line at different $\zeta$'s. The 
full curves moving away from the initial function correspond to
LO results for $Q^2 = 10,\ 10^2,\ 10^{14}\ {\rm GeV}^2$, respectively.
The long-dashed lines give the NLO results for the same values of the 
momentum scale in the same order.}
\label{fig5}
\end{figure}

The factorization proof for diffractive meson production 
\cite{col} opened a
large class of semi-inclusive observables to stringent
QCD analyses \cite{OFPD}. 
The OFPDs (or skewed parton distributuions etc.) provide a new type of
specific information on hadronic wavefunctions. They are of special
importance in connection with internal orbital angular momentum
of e.g. the quarks and gluons in a nucleon \cite{Ji}. It even seems as if 
it must be possible to formulate any observable sensitive to these
internal angular momenta in terms of OFPDs. With planned experiments 
aiming at a determination of the spin-polarized gluon distribution
$\Delta G(x,Q^2)$ one of the main aims of future spin physics
experiments will be to determine the still remaining quantities in the
total angular momentum sum of the nucleon, namely the orbital angular
momentum contributions. Much theoretical work is still needed to
understand their status better. 
As an example figure \ref{fig4} shows a typical example for 
the result of $Q^2$-evolution for the 
angular momentum distributions as defined by us \cite{haeg}.
We find that the large $\Delta G(x)$ generated by $Q^2$-evolution in
NLO is mainly balanced by $L_G(x)$.
While it seems obvious that many
observables should depend on it, it was not possible so far to derive
in a formally clean manner a precise relation between any observable
and the specific correlators associated with orbital angular momnneta. 
The situation is very tricky as the way in which calculations are 
currently done
one has to fix a specific gauge and factorization scheme. 
It is controversal whether a gauge independent formulation can be
found. It might be that the value of those correlators which
correspond 
to our naive
understanding of orbital angular momentum depend on the gauge
considered. This would not be a principle problem (the singlet quark
spin $\Delta \Sigma$ depends e.g. also on the chosen factorization
schmeme), it just needs carefully continued studies.\\
One problem of OFPDs which is basically solved concerns their
$Q^2$-evolution. By now there do exist running LO and NLO codes both
for the singlet and non-singlet channel. Figure \ref{fig5}
shows a typical example for the results of LO- and
NLO-$Q^2$-evolution.
The main message of this figure is that the differences between 
NLO and LO are not large, such that one can expect that also 
the effect of all  higher orders is small.
Studies of the evolution
effects should help to identify those OFPD-models which are relatively
stable under evolution and thus physically acceptable.\\
Another major theoretical problems are to extend the validity of the
factorization proofs to a wider class of reactions, which should be
possible. Also here the work is ongoing. Finally, there is the question
what these proofs imply in praxis. Basically, they show that
the leading contributions in $1/Q^2$ can be parametrized by twist-2
OFPDs.
For practical experiments it is, however, crucial to know the
proportionality factor in front of the $1/Q^2$-power-corrections 
as it determines how large $Q^2$ has to be such
that the higher-twist terms become really negligable. This is basically
still unknown, but first rough estimates give hints that these factors
are disquietingly large \cite{OFHT}. Experimentally, this implies that for the
measurement of OFPDs one nedds a large kinematic region, high
statistics and the
possibility to measure in a clean way the final state hadrons.
Meeting these three requirements is basically the definition of 
an {\sc epic}-type machine. The investigation of OFPDs should
therefore become a major issue in its program.
This topic is covered in far more detail in the contribution  
by Mark Strikman.\\ 

\section{Nuclear Effects}
\begin{figure}[htb]
\centerline{\mbox{\psfig{file=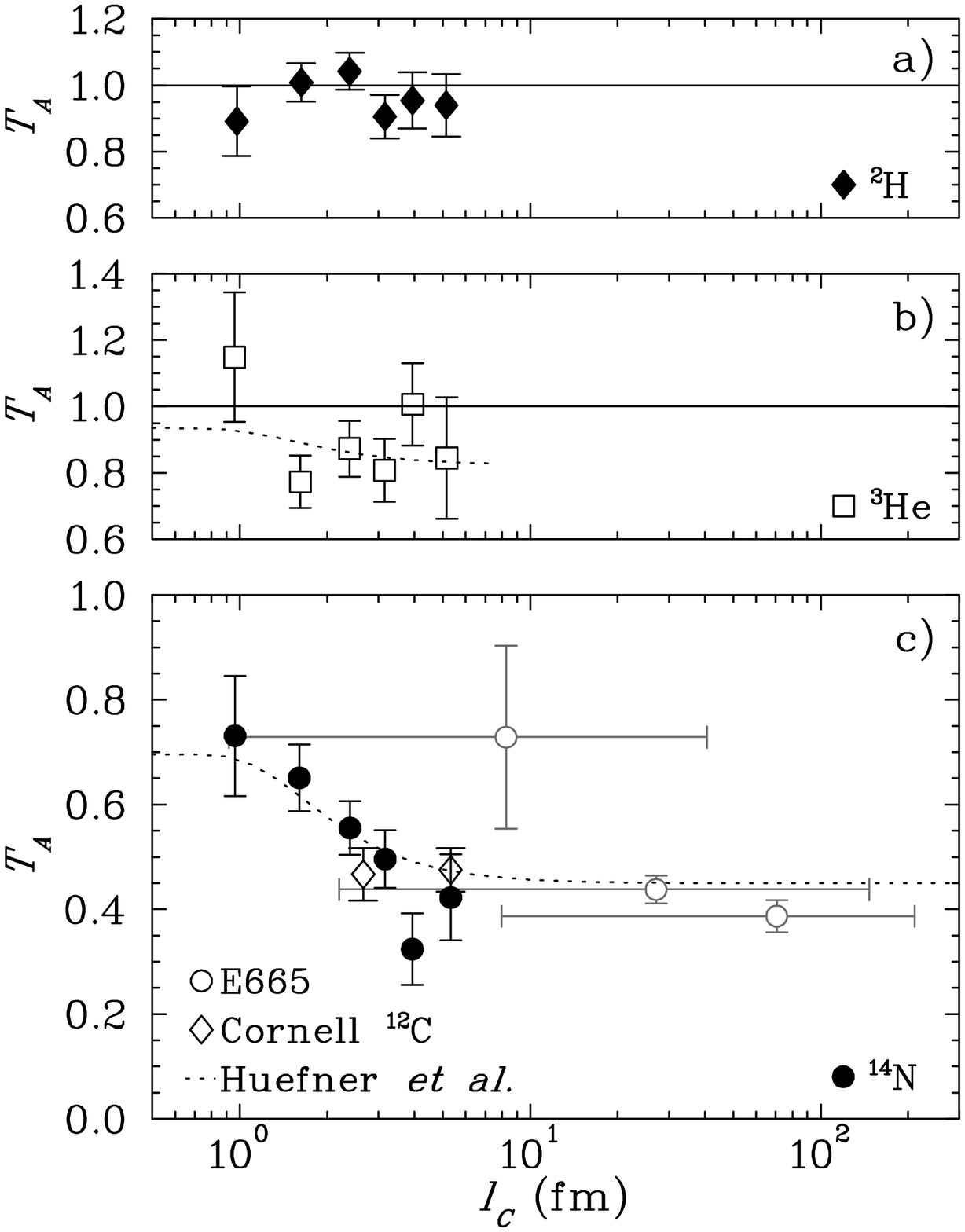,width=0.7\linewidth}}}
\bigskip
\caption{Dependence of the nuclear transparency
  $T_A=\sigma_A/(A\sigma_H)$
as a function of the coherence length $l_c=2\nu/(Q^2+M_{q+\bar q}^2)$.
The dashed curve is the theoretical prediction .}
\label{fig6}
\end{figure}

Diffractive reactions similar to those  just discussed for the case of OFPDs 
play also a major role for e+A collisions. 
First of all their cross-sections are sizeable. In the limit of a
completely black disc, quantum mechanics implies that it is half of the
total cross section. For the values of the invariant energy-squared
$s$ we are considering, a nucleus is not a
black disc, but estimates suggest that diffractive cross sections are
comparable in size to usual ones. The next point is that there is an
intimate connection between shadowing and diffraction. Gribov theory 
unambiguously relates e.g. diffractive processes in the scattering of
a projectile off a single nucleon to nuclear shadowing due to the
interaction of the projectile with two nucleons. Also one should keep
in mind that while it is customary to discuss all nuclear effects in
terms of shadowing and anti-shadowing the underlying microscopic
interaction mechansims are rather varied and there are many additional
detailed questions to be answered. As an example figure \ref{fig6}
shows the
dependence of the nuclear transparency  $T_A$ on the coherence length 
$l_c$ as recently measured 
by {\sc hermes}. One definitely would like to study this effect as a
function of $Q^2$ up to values in which purely perturbative processes
take over, which would require, however, much higher energy. 
This high energy is generally needed to connect smoothly
reactions taking place primarily within a nucleus with those
taking mainly place outside of it. \\
A question of general concern is how large higher-twist effects in
nuclei are in general and whether perturbative QCD still makes sense 
for reactions taking place inside of nuclei. The main issue here are
effects
due to  the strong, delocalized soft gluonfields in nuclei.
Thus the discussion is related to that of gluon saturation 
in nuclei. A powerful approach to deal with this problem was developed
by Sterman and collaborators \cite{ster,guo} (which is too involved to
be discussed here in detail). The paper by Guo \cite{guo} 
illustrate very nicely that the effects are potentially of order 100 percent.
Clearly this issue requires more intensive theoretical and
experimental studies.

\section{Conclusions}
I have briefly sketched some questions to be addressed by an {\sc
  epic}-type
collider. The basic aim of such a machine
would be to explore the {\bf internal structure 
of hadrons} beyond simple twist-2 distribution functions in an infinite
mometum frame (when seen from the high-energy perspective) and beyond
'effective models' (when seen from the traditional nuclear physics 
perspective). The second important aim would be to understand better 
the {\bf QCD dynamics} of reactions taking place in nuclei.
Keeping in mind that there was tremendous theoretical
progress in both fields during the last few years 
it is clear that with improved understanding more and  more 
observables will become accessible to a detailed interpretation.
Extrapolating the current development, it is therefore obvious that in
a few years from now a still much larger collection of 
interesting clearly interpretable semi-inclusive
and exclusive electron-nucleon/nucleus reactions would wait to be 
investigated experimentally. Basically all of these would require,
however, an {\sc epic}-type machine.

\end{document}